\documentclass[12pt]{article}
\usepackage{authblk}
\usepackage[a4paper, left=1cm, right=1cm, top=1.5cm, bottom=1.5cm]{geometry}
\usepackage{mhchem}
\usepackage{caption}
\usepackage{xparse}
\usepackage{multicol} 
\usepackage{multirow}

\usepackage{adjustbox}
\usepackage{graphicx}
\usepackage{float}
\usepackage{array}
\usepackage{ragged2e}

\makeatletter
\def\ttabular{%
\hbox\bgroup
\let\\\cr
\def\rulea{\ifnum\rowc=\@ne \hrule height 1.3pt \fi}
\def\ruleb{
\ifnum\rowc=1\hrule height 1.3pt \else
\ifnum\rowc=6\hrule height \heavyrulewidth 
   \else \hrule height \lightrulewidth\fi\fi}
\valign\bgroup
\global\rowc\@ne
\rulea
\hbox to 10em{\strut \hfill##\hfill}%
\ruleb
&&%
\global\advance\rowc\@ne
\hbox to 10em{\strut\hfill##\hfill}%
\ruleb
\cr}
\def\endttabular{%
\crcr\egroup\egroup}

\usepackage[numbers,sort&compress,square,super]{natbib}
\bibpunct{[}{]}{,}{n}{\textsuperscript{}}{}
\newcommand{\supercite}[1]{\textsuperscript{\cite{#1}}}

\begin{document}
\title{
\LARGE Experimental evidence of crystal-field, Zeeman splitting, and spin-phonon excitations in the quantum supersolid \ce{Na_{2}BaCo(PO_{4})_{2}}}
\author{\large Ghulam Hussain\textit{$^{1\ddag}$}, Jianbo Zhang\textit{$^{2\ddag}$}, Man Zhang\textit{$^{1}$}, Lalit Yadav\textit{$^{3}$}, Yang Ding\textit{$^{2}$}, Changcheng Zheng\textit{$^{1}$}, Sara Haravifard\textit{$^{3}$} and Xiawa Wang\textit{$^{\ast}$}}
\affil[1]{\normalsize Division of Natural and Applied Sciences, Duke Kunshan University, Kunshan, Jiangsu 215316, China}
\affil[2]{\normalsize Center for High-Pressure Science and Technology Advanced Research, Beijing 100094, China}
\affil[3]{\normalsize Department of Mechanical Engineering and Materials Science, Duke University, Durham, NC 27708, USA}
\affil[${\ddag}$]{These authors contributed equally to this work.}
\affil[${\ast}$]{Corresponding Author: xiawa.wang@dukekunshan.edu.cn}
\date{}
\maketitle


\begin{abstract}
\justifying
Drawing inspiration from the recent breakthroughs in the \ce{Na_{2}BaCo(PO_{4})_{2}} quantum magnet, renowned for its spin supersolidity phase and its potential for revolutionary cooling applications, our study delves into the intricate interplay among lattice, spin, and orbital degrees of freedom within this intriguing compound. Using meticulous temperature, field, and pressure-dependent Raman scattering techniques, we present compelling experimental evidence revealing pronounced crystal-electric field (CEF) excitations, alongside the interplay of CEF-phonon interactions. Notably, our experiments elucidate all electronic transitions from $j_{1 / 2}$ to $j_{3 / 2}$ and from $j_{1 / 2}$ to $j_{5 / 2}$, with energy level patterns closely aligned with theoretical predictions based on point-charge models. Furthermore, the application of a magnetic field and pressure reveals Zeeman splittings characterized by Land\'e-g factors as well as the CEF-phonon resonances. The anomalous shift in coupled peak at low temperatures originates from the hybridization of CEF and phonon excitations due to their close energy proximity. These findings constitute a significant step towards unraveling the fundamental properties of this exotic quantum material for future research in fundamental physics or engineering application.

\vspace{5mm}
\noindent\textbf{Keywords:} \ce{Na2BaCo(PO4)2}, crystal-field splitting, Raman spectroscopy, cobalt compound, spin-phonon interactions

\end{abstract}

\vspace{55mm}
\section{Introduction}
\hspace{1em} In contemporary condensed matter physics, a significant endeavor revolves around identifying an isotropic and geometrically frustrated system capable of hosting an ideal two-dimensional (2D) triangular lattice. This system should be characterized by effective spin-1/2 local moments, devoid of structural imperfections or inherent chemical disorder.\supercite{peri2024probing,tian2024dimensionality,sarkar2019quantum,bordelon2019field,ding2019gapless} \ce{Na_{2}BaCo(PO_{4})_{2}} (NBCP) has emerged as one of the most compelling candidates due to its nearly perfect triangular lattice structure devoid of inherent chemical disorder or site-mixing.\supercite{zhong2019strong} From previous experimental results, the material not only unveiled intriguing emergent quantum phenomena, including spin-orbit coupling (SOC), crystalline field effects, topological order, and fractionalized excitations deviating from conventional spin wave behaviors,\supercite{chen2015fractionalization,castelnovo2012spin,sheng2024continuum,sheng2022two} but also offered promising avenues for next-generation advancements in energy storage, conversion, magneto-caloric effects, and novel computational paradigms.\supercite{xiang2024giant,liu2022quantum,broholm2016basic} Experimental evidence supports NBCP as a host for a spin supersolid phase,\supercite{xiang2024giant,gao2022spin} with \ce{Co^{2+}} ions exhibiting effective spin-1/2 characteristics. Moreover, NBCP displays a broad continuum observed in inelastic neutron scattering, attributed to spinon excitations, alongside a significant antiferromagnetic exchange without long-range ordering down to 0.05 K.\supercite{sheng2022two,zhong2019strong} Subsequent investigations via ultra-low temperature specific heat and thermal conductivity measurements indicated an antiferromagnetic phase transition occurring at the N\'eel temperature ($T_\mathrm{N}$) of 148 mK.\supercite{li2020possible,huang2022thermal} Recent findings from muon spin relaxation ($\mu$SR) and nuclear magnetic resonance experiments suggest magnon-like magnetic excitations transitioning to spinons at the critical field $\mu_0H_\mathrm{C}$ = 1.65 T.\supercite{lee2021temporal}The exotic properties exhibited by NBCP provide valuable insights for advancing material research, with the potential to revolutionize existing paradigms and enable the development of materials surpassing current limits. Hence, a comprehensive exploration of NBCP properties is imperative for its thorough understanding and subsequent development.
\par
\vspace{5mm}
\noindent
In this study, we present a comprehensive analysis of the intricate magnetic interactions within NBCP, employing temperature, polarization, magnetic field, and pressure-dependent Raman spectroscopy down to 7 K with magnetic field up to $8 \mathrm{~T}$ and pressure up to 2.18 GPa. Our investigations reveal all the primary phonons and crystal-field excitations (CEF) with their energies corroborated through density-functional-theory (DFT) and point-charge model calculations. In addition, we discovered peak splittings with the application of magnetic field and pressure, indicating different mechanisms that originated from Zeeman effect, CEF-phonon coupling, and anisotropy of the Land\'e-g factor. This work constitutes the first direct experimental evidence of the intricate particle interplays in NBCP under extreme conditions, thereby opening new avenues for exploring its exotic properties and facilitating further research in this promising material.

\section{Experimental Methods}
\hspace{1em} We synthesized high-quality \ce{Na_{2}BaCo(PO_{4})_{2}} single crystals using the conventional flux method.\supercite{zhong2019strong} Initially, polycrystalline NBCP samples were prepared via solid-state reaction. The precursor powders, including \ce{Na_{2}CO_{3}} (99\%, Alfa Aesar), \ce{BaCO_{3}} (99.99\%, Adamas), \ce{CoO} (99\%, Alfa Aesar), and \ce{(NH_{4})_{2}HPO_{4}} (99.5\%, Sigma-Aldrich), were thoroughly mixed using an agate mortar and pestle. The mixture was then annealed in air at 700 $^\circ$C for 24 hours. Subsequently, the dried powders were combined with $\mathrm{NaCl}$ flux (in a molar ratio of 1:5) and placed in a platinum crucible covered with a lid inside a box-type furnace. The crucible was heated to 950 $^\circ$C, held for 2 hours for homogenization, and slowly cooled to 750 $^\circ$C at a rate of 2 $^\circ$C/hour. Pink crystals were obtained, washed in water, and manually separated from the bulk, resulting in typical dimensions of $2 \times 1 \times 0.3 \mathrm{~mm}^3$, as depicted in the Figure 1 b(inset). Raman spectroscopy was performed using linearly polarized lasers with wavelengths of 488 nm and 532 nm, yielding a spot size of approximately 5 $\mu$m on the sample employing a grating of $1800 \mathrm{~g/mm}$. Spectra were collected in parallel (XX) and crossed (XY) polarization configurations, with phonon vibrational anisotropy explored by rotating the laser polarization direction. For temperature and field-dependent measurements, single crystals were placed in a helium closed-cycle cryostat, allowing temperature variation between 7 and 300 K. Crystal orientations in the ab and ac planes were accessed by placing samples on the cold stage in two directions. Temperature and field-dependent spectra were recorded in both planes, with an exposure time of 90 s and a laser power reaching the cryostat's optical window of approximately 8 mW for each spectrum.
\par
\vspace{5mm}
\noindent
The calculation of zone-centered phonons was done using the plane-wave approach implemented in Quantum Espresso.\supercite{hinuma2017band,perdew1996generalized} All the calculations were carried out using Optimized Norm-Conserving Vanderbilt pseudo-potentials. The plane-wave cutoff energy was set to 120 Ry. Dyanmical matrix and eigenvectors were determined using density functional perturbation theory. The numerical integration over the Brillouin zone were done with a 191 k-point mesh in the Monkhorst-Pack grid. In our calculations, we use fully relaxed ionic positions with experimental lattice parameters given in Ref.\supercite{zhong2019strong}

\section{Results and Discussions}
\subsection{\textbf{Raman spectroscopy and selection rules}}
\hspace{1em} \ce{Na_{2}BaCo(PO_{4})_{2}} crystallizes in a trigonal structure with lattice parameters $\alpha=\beta=90^{\circ}, \gamma=120^{\circ}$, and lattice constants $a = b = 5.31275$ Å, $c = 7.0081$ Å in space group $\mathrm{P} \overline{3} \mathrm{m} 1$ (No. 164) and point group $D_{3 d}^3$ ($\overline{3} \mathrm{m} 1$ in Hermann-Mauguin notation) as depicted in Figure 1(a-b). The triangular layer of magnetic \ce{CoO_{6}} octahedra resides in the ab plane with each  \ce{Co^{2+}} ion coordinated to six oxygen atoms. With a layered AA-stacking along the c-axis, the \ce{Co^{2+}} ions form an effective two-dimensional (2D) triangular lattice ideally to be free of chemical disorders or structural distortions. 
\begin{figure}[H]
       \centering
       \includegraphics[width=.74\textwidth]{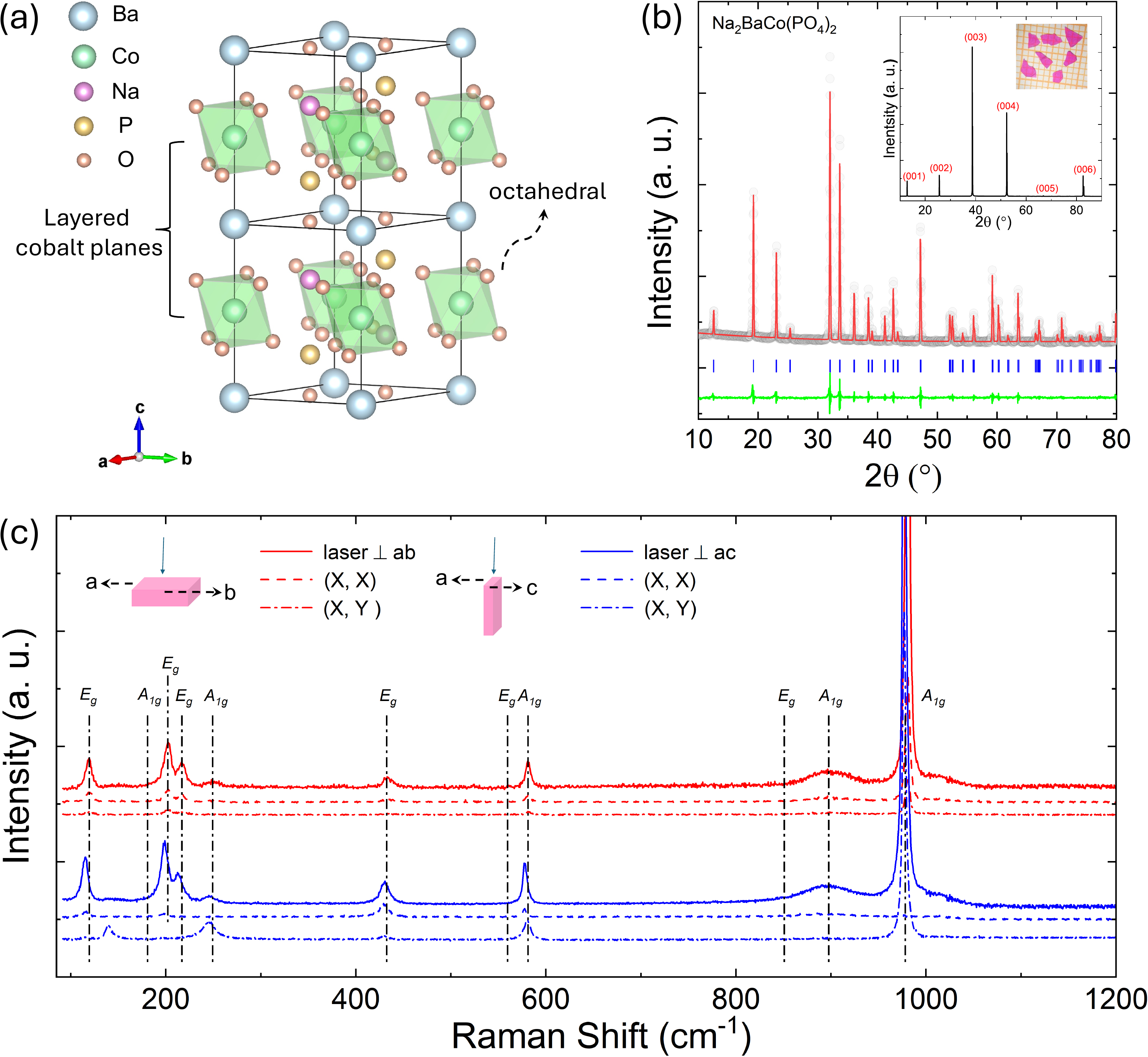}
       \caption{(a) NBCP crystal structure. 
 (b) The room temperature powder XRD pattern and inset showing the single crystal XRD, indicating the well-grown (001) surface with a photograph image. (c) Spectra were collected in parallel (XX) and crossed (XY) polarization configurations on the ab and ac planes of the NBCP single crystals at room temperature.}
\end{figure}

\noindent
Polarization-dependent Raman spectra at room temperature under ambient pressure are depicted in Figure 1(c). The bulk unit cell comprises a total of 14 atoms, resulting in 42 normal phonon modes at the $\Gamma$ point. The \ce{Co} and \ce{Ba} atoms occupy \ce{C_{3v}} sites in each unit cell, while the remaining 6 \ce{O} atoms binding to $\mathrm{Co}$ atoms occupy the \ce{C_{s}} sites. The vibrational modes at the $\Gamma$ point can be categorized into $\Gamma (D_{3 d}^3) = 5 A_{1 g} + 7 A_{2 u} + 6 E_g + 8 E_u + A_{2 g} + A_{1 u}$. Among these vibrational modes, 11 are Raman active modes corresponding to the non-degenerate $A_{1g}$ symmetry modes and the doubly degenerate $E_g$ symmetry modes. Moreover, the energy values between experimental values and ab-initio computations by Quantum Espresso matched very well as seen in the supplementary Table S1.\supercite{zhong2019strong,soyalp2013ab,ye2023calculating,abragam2012electron} The symmetry analysis reveals that the $E_g$ mode can be observed in both cross and parallel configurations, while the $A_{1g}$ mode is visible only in the parallel polarization configuration as verified by the experimental observation in supplementary Figure S1. The Raman tensors of the two modes are given as shown in Equation (1)-(3).\supercite{bordelon2019field}
\begin{align}
& R\left(A_{1 g}\right)=\left(\begin{array}{lll}
a & \cdot & \cdot \\
\cdot & a & \cdot \\
\cdot & \cdot & b
\end{array}\right) \\
& R\left(E_{1 g}\right)=\left(\begin{array}{ccc}
c & \cdot & \cdot \\
\cdot & -c & d \\
\cdot & d & \cdot
\end{array}\right) \\
& R\left(E_{1 g}\right)=\left(\begin{array}{ccc}
\cdot & -c & -d \\
-c & \cdot & \cdot \\
-d & \cdot & \cdot
\end{array}\right)
\end{align}
The intensity of a Raman mode, denoted as $I$, is proportional to $\mid e_i \cdot R \cdot e_s \mid^2$, where $R$ is the Raman tensor of the phonon mode, and $e_i$ and $e_s$ are the unitary polarization vectors of the incident and scattered light, respectively.\supercite{alencar2020raman,klein2005raman} For the $A_{1 g}$ mode, by substituting the Raman tensor, we can obtain the mode's different responses to polarization as shown in Equation (4) - (5). Under the parallel configuration where $\gamma=\theta$, the intensity $I(A_{1g})$ is proportional to $|a|^2$, resulting in a circular polarization profile. Conversely, under the experimental cross configuration where $\gamma=\theta+90^\circ$, $I(A_{1g})=0$, indicating that the mode is non-visible as confirmed by the experiment. Similarly, for the $E_g$ mode, we have $I(E_g) \propto |c|^2$ so that the polarization profile is circular and visible for both polarizations.
\begin{align}
& I\left(A_{1 g}\right) \propto\left|(\cos \gamma, \sin \gamma, 0) \cdot\left(\begin{array}{ccc}
a & \cdot & \cdot \\
\cdot & a & \cdot \\
\cdot & \cdot & b
\end{array}\right) \cdot\left(\begin{array}{c}
\cos \theta \\
\sin \theta \\
0
\end{array}\right)\right|^2 \\
& I\left(A_{1 g}\right) \propto|a|^2 \cos ^2(\gamma-\theta)
\end{align}

\subsection{\textbf{Crystal field excitations}}
\hspace{1em} We characterized the temperature-dependent (7 - 300 K) Raman spectra of NBCP with the whole set of data given in supplementary Figure S2(a–b). Below 130 K at ambient pressure, additional five peaks emerged. These peaks appeared at 309 $\mathrm{cm}^{-1}$ (38 meV), 346 $\mathrm{cm}^{-1}$ (42 meV), 862 $\mathrm{cm}^{-1}$ (107 meV), 900 $\mathrm{cm}^{-1}$ (112 meV), and 1036 $\mathrm{cm}^{-1}$ (128 meV) demonstrating strong temperature-dependent intensities, weak polarization dependence, and linewidth softening as the temperature decreased. The observed energy values can be roughly grouped into two transition levels: two peaks around 40 meV and another three around 110 meV. The two lower energy peaks shown in Figure 2(b-c) around 40 meV persisted roughly to $130 \mathrm{~K}$ and demonstrated opposite shifting directions as phonons as shown in Figure 2(d). Two higher energy peaks with energies at 107 meV and 112 meV showed up as one broad peak persisting all the way to room temperature as shown in Figure 2(e). Upon deconvolution, we discovered four peaks inside, which were the two phonon peaks of $E_{g}$ at 868 $\mathrm{cm}^{-1}$, $A_{1g}$ at 886 $\mathrm{cm}^{-1}$ together with the two new peaks at 862 $\mathrm{cm}^{-1}$ (107 meV) and 900 $\mathrm{cm}^{-1}$ (112 meV) as shown in Figure 2(f) and supplementary Figure S2(c-d).  Figure 2(g) showed the anomalous shifting directions for these peaks, suggesting potential hybridization or couplings that will be discussed later. A third newly emerged peak at 1036 $\mathrm{cm}^{-1}$ (128 meV) was further away from any phonon peaks and appeared as an independent one below 130 K.  
      \begin{figure}[H]
       \centering
       \includegraphics[width=1.\textwidth]{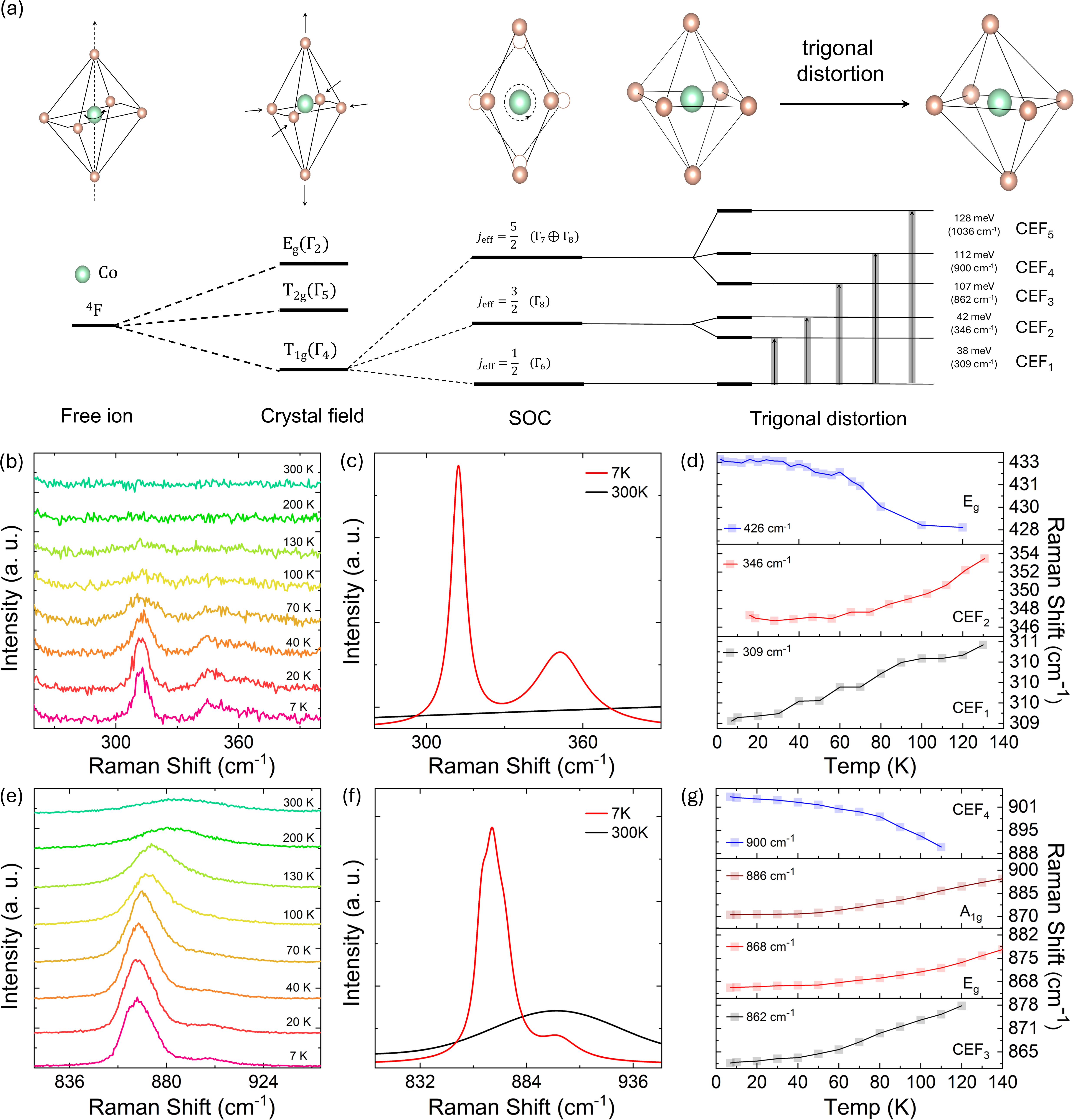}
       \caption{(a) The energy levels of the ground $^4F$ \ce{Co^{2+}}split by the crystal field in the presence of SOC and trigonal distortion. (b) The temperature-dependent CEF excitation peaks around 40 meV. (c) A comparison of low and room temperature Raman spectra the 40 meV CEF peaks. (d) Peak position evolution as a function of temperature for phonon and 40 meV CEF peaks. (e) The temperature-dependent CEF excitation peaks around 110 meV. (f) A comparison of low and room temperature Raman spectra for the 110 meV CEF peaks (two lower peaks). (g) Peak position evolution as a function of temperature for the two lower 110 meV CEF peaks and adjacent phonon peaks. }
      \end{figure}
\noindent
We attribute these additional peaks to crystal-field excitations in the presence of spin-orbit coupling and small trigonal distortion and further verified our conclusion by the point-charge model computation and theoretical values reported in literature. In \ce{Na_{2}BaCo(PO_{4})_{2}}, the cobalt ion exists in the \ce{Co^{2+}} state with an electronic configuration of $3d^7$. Figure 2(a) illustrates the energy diagram for the cobalt ion situated within an octahedral crystal field under the influence of spin-orbit coupling. Expressed in term symbols, the electronic ground state for the \ce{Co^{2+}} ion is $^4F$ with $\mathrm{L}=3$ and $\mathrm{S}=3/2$, resulting in $2\mathrm{L}+1=7$ initially degenerate levels for the free ion. Subjecting to the octahedral field, the energy levels split into the $T_{1g}$ mode (subject to $\Gamma_4$ symmetry), the $T_{2g}$ mode ($\Gamma_5$), and higher $A_{2g}$ mode ($\Gamma_2$). Spin-orbit coupling further splits the energy levels into a $j=1/2$ doublet (subject to $\Gamma_6$ symmetry), a $j=3/2$ quartet ($\Gamma_8$), and a $j=5/2$ sextet with four transforming according to $\Gamma_7$ and the others according to $\Gamma_8$ symmetries. The experimental results also agreed with a recent theoretical paper on the computation of NBCP's Raman spectroscopy with a small negative trigonal distortion so that ligands are pushed closer together in the xy-plane while being pulled farther apart along the z-axis, resulting in a weak $j=5/2$ peak plus two stronger $j=5/2$ peaks at higher wavenumbers.\supercite{mou2024comparative} In the low-temperature measurement without magnetic field, the Hamiltonian can be represented according to Equation (6).
\begin{equation}
H  =  H_{\text {Crys}} +H_{\text {SOC}} \\
  =  \sum_{n, m} B_n^m O_n^m + \lambda_{SO} S . L \\
\end{equation}
Here, $H_{\text{Crys}}$ represents the crystal field contribution, which can further be expanded into a series of spherical harmonics where $B_n^m$ are the CEF multiplicative factors related to the strength of the crystal field and $O_n^m$ belong to the Stevens' operators. The spin-orbit coupling $H_{\text{SOC}}$ can be expressed as the dot product of the operators with a spin-orbit coupling constant $\lambda_{SO}$. By directly constructing a point charge model based on the NBCP structure, the calculations were performed using the open-source package PyCrystalField. Table 1 compares our measured energy levels with PyCrystalField computation \supercite{scheie2021pycrystalfield} and the theoretical values done by density-functional-theory.\supercite{wellm2021frustration} The computed CEF constants were $B_2^0=-3.75$ meV, $B_4^0=0.44$ meV, and $B_4^3= 11.50$ meV, with the spin-orbit coupling constant $\lambda_{SO} = -22$ meV from literature.\supercite{abragam2012electron}

\begin{table}[htbp]
    \centering
    \caption{Comparison of experimental and calculated CEF levels}
    \begin{tabular}{>{\centering\arraybackslash}m{5cm}>{\centering\arraybackslash}m{5cm}>{\centering\arraybackslash}m{5cm}}
        \hline
        Experimental observed & Point charge calculation & DFT computation\supercite{wellm2021frustration} \\
        \hline
        $38\,\text{meV}\,(309\,\mathrm{cm}^{-1})$ & $37.3\,\text{meV}$ & $40.7\,\text{meV}$ \\
        $42\,\text{meV}\,(346\,\mathrm{cm}^{-1})$ & $43.2\,\text{meV}$ & $45.6\,\text{meV}$\\
        $107\,\text{meV}\,(862\,\mathrm{cm}^{-1})$ & $104.2\,\text{meV}$ & $112.4\,\text{meV}$\\
        $112\,\text{meV}\,(900\,\mathrm{cm}^{-1})$ & $113.4\,\text{meV}$ & $113.6\,\text{meV}$\\
        $128\,\text{meV}\,(1036\,\mathrm{cm}^{-1})$ & $133.4\,\text{meV}$ & $128.3\,\text{meV}$\\
        \hline
    \end{tabular}
\end{table}
\par
\noindent
The peak positions are in excellent agreement with the transition energies from the ground state $j=1/2$ to the excited $j=3/2$ and $j=5/2$ states. The two lower-energy peaks arose from the smaller energy gap between $j=1/2$ and $j=3/2$, while the three higher energy ones corresponding to the transition from $j=1/2$ to $j=5/2$. These transitions involved a change in the angular momentum quantum number, leading to a non-zero transition moment that renders them observable in Raman spectroscopy. Our results are also consistent with another recent report for this material indicating the presence of unconventional magnetism driven by spin-orbit coupling, electronic correlations, and CEF excitations.\supercite{mou2024comparative} 

\subsection{\textbf{Zeeman splitting and CEF-phonon coupling}} 
\hspace{1em} In crystal field theory, metal–ligand interactions are typically considered to be solely electrostatic, suggesting field-independent Raman excitations at low temperatures. However, the behaviors of the CEF peaks were more complex in NBCP as shown in the supplementary Figure S4(a-b), revealing intricate particle interplays inside. For the lower energy peaks that we designated as transition from  $j=1/2$ and $j=3/2$, the two peaks exhibited shifts to higher frequencies at a rate of $\sim 1.7 \mathrm{~cm}^{-1} / \mathrm{T}$ and $\sim 1.0 \mathrm{~cm}^{-1} / \mathrm{T}$, respectively. Moreover, when the field strength increased beyond 3 T, a third peak moving in the opposite direction appeared, which we attributed to Zeeman splitting, as illustrated in Figure 3(a). To validate our conclusion, we applied magnetic field in both parallel and perpendicular directions. When a magnetic field is applied, the peaks that correspond to opposite angular momenta move away from each other, resulting in the Zeeman phenomenon. In our NBCP sample, the 38 meV peak showed \(\Delta E \sim 14 \mathrm{~cm}^{-1}\) in response to the 8 T applied field in both H$\parallel$c and H$\parallel$a directions as shown in Figure 3(b). When a pressure of P = 2.18 GPa was applied, the peak splitting width increased from \(\Delta E \sim 14 \mathrm{~cm}^{-1}\) to \(\Delta E \sim 22 \mathrm{~cm}^{-1}\) as shown in Figure 3(c), signifying that the energy levels are more distinct under pressure. 
\par
\vspace{5mm}
\noindent
Typically, the interaction between an external magnetic field and the magnetic dipole moment associated with the total angular momentum of an ion leads to Zeeman splitting. We used the energy difference between the split modes to calculate the effective magnetic moment ($\mu_{\mathrm{eff}}$) and the associated $g$-factor using the equation $\Delta E = 2 g \mu_\mathrm{B}B$, where $\Delta E$ represents the peak-peak splitting for a given applied magnetic field $B$, and $\mu_\mathrm{B}$ denotes the Bohr magneton. The extracted data as a function of magnetic field with a linear fitting are shown in Figure 3(d). A g-factor of $g_{\perp}=4.34$ and $g_{\parallel}=4.80$ values and the effective magnetic moment $\mu_{\mathrm{eff}} \approx 1.14 \mu_\mathrm{B}$ can be extracted as seen in Figure 3(d, e) that is in excellent agreement with the Electron Spin Resonance (ESR) experiment values of $g_{\perp}=4.22$ and $g_{\parallel}=4.81$.\supercite{zhong2019strong, wellm2021frustration} Concurrently, the other peak with a similar transition energy of 42 meV exhibited broadening without splitting at all. For \ce{Na_{2}BaCo(PO_{4})_{2}}, the $g$-factor was theoretically predicted to be anisotropic with the small trigonal distortion to be one of contributors.\supercite{gao2022spin, winter2022magnetic} It is plausible that the different transition observations in the 38 meV and 42 meV peaks were caused by the anisotropy of the $g$ factor, resulting from different angular momenta change to show up under magnetic field.\supercite{zhong2019strong,lee2018spin,wellm2021frustration}

     \begin{figure}[H]
       \centering
       \includegraphics[width=1.\textwidth]{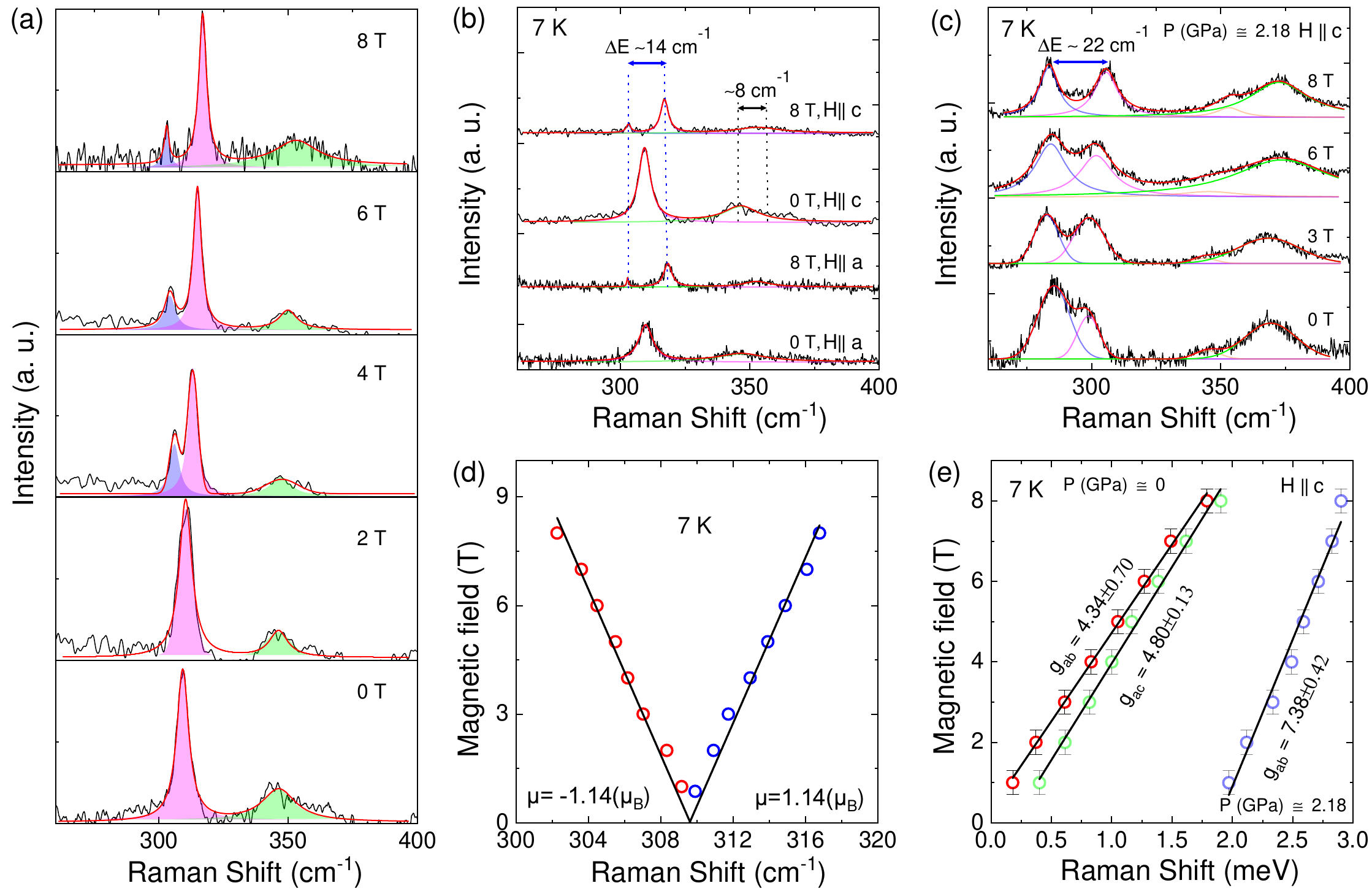}
       \caption{(a) Zeeman splitting of 34 meV peak at lower temperature (7 K) with increasing magnetic field perpendicular to the ab plane. (b) The comparison of the spectra with and without magnetic field in different orientations. (c) Under P= 2.18 GPa applied pressure, the 40 meV peaks evolves as a function of magnetic field. (d) The Zeeman splitting valley showing the  peak shift as a function of the field. (e) The slope fit yielded $g$ values obtained from the energy shift as a function of the applied magnetic field.} 
       \end{figure}
The higher energy Raman mode, centered around 110 meV, responded to the magnetic field in a markedly different way. According to the point-charge model, there are three transitions arising from $j = 1/2$ to $j = 5/2$ located at 862 $\mathrm{~cm}^{-1}$, 900 $\mathrm{~cm}^{-1}$ and 1036 $\mathrm{~cm}^{-1}$. However, within a similar energy range, two phonons: the $E_g$ mode at 863 $\mathrm{~cm}^{-1}$ and the $A_{1g}$ mode at 870 $\mathrm{~cm}^{-1}$ are also present. These two modes come from the \ce{PO_{4}} tetrahedron' symmetric stretching and asymmetric stretching vibration modes for the corresponding phosphate with the glasrite-type phase.\supercite{nakamoto2009infrared,lee2008raman,zhai2011high} This suggests a potential coupling between the CEF and phonon modes contributing to this peak. Typically, the electronic excitations that are close to phonons can come from the CEF and the magnetic ions' spin-orbit effects, where the spin and orbital degree of freedom can interact strongly to magnetic field. To investigate this hypothesis, we deconvoluted the peak into four contributors in the spectral range $800-980 \mathrm{~cm}^{-1}$, among which two are CEF peaks and two are phonon peaks, and analyzed their frequency dependence on the field strength as shown in Figure 4(a-b). In the absence of magnetic field, the lowest CEF peak at 107 meV ($862 \mathrm{~cm}^{-1}$) showed the smallest intensity buried by the two phonons. When field strength increased, the phonons displayed minimal position shift, but showed a strong intensity exchange with the adjacent CEF mode up to $8 \mathrm{~T}$ along both the ab and ac plane. Notably, the CEF peak at $862 \mathrm{~cm}^{-1}$ (107 meV) showed a modest shift rate of approximately $0.30 \mathrm{~cm}^{-1} / \mathrm{T}$ below $3 \mathrm{~T}$, escalating to $1 \mathrm{~cm}^{-1} / \mathrm{T}$ above $3 \mathrm{~T}$. Furthermore, the linewidth and intensity of the CEF mode at $862 \mathrm{~cm}^{-1}$ (107 meV) exhibited stronger field dependence compared to $900 \mathrm{~cm}^{-1}$ (112 meV) CEF mode as shown in Figure 4(a). A cross-over between the $E_g$ phonon and the lowest CEF peak at 107 meV occurred around 3T with a sharp intensity exchange between the two modes. With continuously increased magnetic field, $E_g$ mode intensity greatly enhanced at the expense of the 107 meV CEF peak, while the total peak intensity remained approximately conserved.
\par
\vspace{5mm}
\noindent
When applying the field in different crystallographic directions, there were obvious differences for this coupling behavior as shown in Figure 4(b). Firstly, only two peaks can be deconvoluted because the $A_{1g}$ phonon mode was only visible in one polarization. However, CEF-phonon energy cross-overs and intensity trade-offs can be seen regardless of the field direction. An applied pressure can easily separate the contributors to this peak as seen in Figure 4(c). When a pressure of 2.18 GPa was applied, the peaks broadened and were separated even without magnetic field. The change of the Raman spectra can possibly be explained by the direction of the trigonal distortion. With a pressure applied in the perpendicular direction, the ligands were pushed closer in the z-axis resulting in a positive distortion instead of the negative one that was intrinsic to non-pressured sample. A positive distortion will result in two stronger $j = 5/2$ CEF peaks as can obviously be seen in the blue and green fittings in Figure 4(c). With a continuously increasing field, the peak separation became more prominent. At the maximum field strength of 8 T, the pressure resolved the peaks and separated them into four distinct contributors. However, a noteworthy point is that CEF-phonon coupling was not present in the pressured spectra. The peak separation was quite significant so that their energies no longer possessed a cross-over. The peak positions as a function of the applied magnetic field with and without pressure are shown in Figure 4(d-e). With applied pressure, the phonon peaks' responsiveness to the magnetic field roughly remained the same level while the CEFs' responsiveness reduced. This can possibly be explained as a more pronounced impact by the lattice distortion on the particle dynamics so that magnetic contributions were diminished comparatively. 

     \begin{figure}[H]
       \centering
       \includegraphics[width=1.\textwidth]{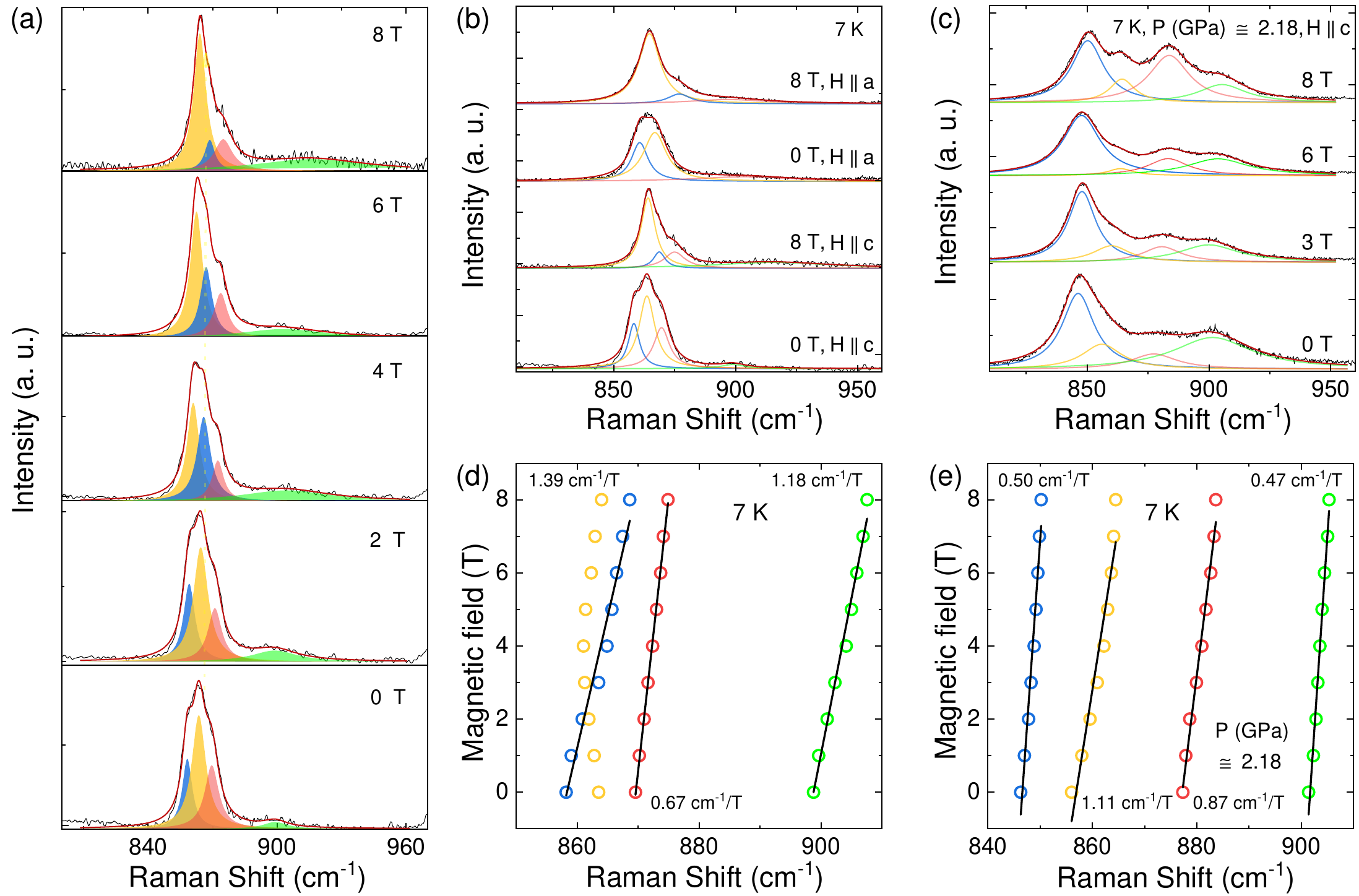}
       \caption{ (a)Deconvolution of Raman spectra $800-960 \mathrm{~cm}^{-1}$. (b) The peak intensity in  response to the applied field along different crystal's axis. (c) Low temperature (T = 7 K) Raman spectra of NBCP at P = 2.18 GPa with an applied magnetic field. Four distinct peaks showed up. (d) Linear fits of the peak shifts as a function of field strength at ambient pressure. (e) Linear fits of the peak shifts as a function of field strength at 2.18 GPa pressure.}
\end{figure}

\section{Conclusion}
\hspace{1em} In summary, we comprehensively investigated the spin-orbit coupling, crystal electric field (CEF) excitations, CEF-phonon interactions in the exotic supersolid compound \ce{Na_{2}BaCo(PO_{4})_{2}} using temperature, magnetic field and pressure-dependent Raman scattering techniques. We successfully identified and confirmed the symmetries and all primary phonon modes in the material. Our findings revealed all the energies of CEF excitations from the ground-state to the excited states, which correlated to the peaks observed at 38, 42, 107, 112, and 128 meV and in excellent agreement with the point-charge model. Additionally, we observed that the peaks at 38 meV and 42 meV which represented the first CEF levels, exhibited field-dependent behavior and Zeeman splitting under magnetic field. The higher energy CEF modes at 107 and 112 meV created a resonance with two ligand phonons which can be easily separated by an applied pressure. The distinct energies associated with spin-orbit coupling, CEF-phonon interactions, and electronic excitations firmly establish NBCP as a promising candidate for further theoretical and experimental investigations into quantum magnet with nearly ideal triangular lattice. Our results shed light on the intricate interactions within NBCP, offering new avenues for this material for further research in both fundamental physics and practical applications.

%
\vspace{15mm}
\bibliography{MSP-template}
\bibliographystyle{MSP}

\end{document}